\documentclass[fleqn,usenatbib]{mnras}
\usepackage{newtxtext,newtxmath}
\usepackage[T1]{fontenc}
\usepackage{ae,aecompl}
\usepackage{amsmath}	% Advanced maths commands
\usepackage{amssymb}	% Extra maths symbols
\usepackage{soul}
\usepackage[english]{babel}
\usepackage{journals}
\usepackage[pdftex]{graphicx,color} 
\usepackage[latin1]{inputenc}
\usepackage{graphics} 
\usepackage{amsfonts} 
\usepackage{multicol} 
\usepackage{layout} 
\usepackage{layouts}
\usepackage{multirow}
\usepackage{float}
\usepackage{tabularx}
\usepackage{siunitx}
\usepackage{subcaption}

\def\msun{{\rm M}_{\sun}}

\title[Illustris subhaloes]{
Dark matter stripping in galaxy clusters: a look at the Stellar to Halo Mass relation in the Illustris simulation
}

\author[Niemiec et al.]
	{\parbox{\textwidth}{
	Anna Niemiec$^{1,2}$\thanks{E-mail: \href{mailto:annaniem@umich.edu} {annaniem@umich.edu}}, 
	Eric Jullo$^{2}$, 
    Carlo Giocoli$^{3,4,5,6}$,
	Marceau Limousin$^{2}$,
    Mathilde Jauzac$^{7,8,9}$
	\\ \\ 
	}
	\vspace*{2pt}\\
$^{1}$Department of Astronomy, University of Michigan, 1085 South University Ave, Ann Arbor, MI 48109, USA\\
$^{2}$Aix Marseille Univ, CNRS, CNES, LAM, Marseille, France \\
$^{3}$Dipartimento di Fisica e Scienze della Terra, Universit\`a degli Studi di Ferrara, via Saragat 1, I-44122 Ferrara, Italy\\
$^{4}$INAF - Osservatorio di Astrofisica e Scienza dello Spazio di Bologna,  via Gobetti 93/3, I-40129 Bologna, Italy \\
$^{5}$Dipartimento di Fisica e Astronomia, Alma Mater Studiorum Universit\`{a} di Bologna, via Gobetti 93/2, I-40129 Bologna, Italy\\
$^{6}$INFN - Sezione di Bologna, viale Berti Pichat 6/2, I-40127 Bologna, Italy\\
$^{7}$Centre for Extragalactic Astronomy, Department of Physics, Durham University, Durham DH1 3LE, UK\\
$^{8}$Institute for Computational Cosmology, Durham University, South Road, Durham DH1 3LE, UK\\
$^{9}$Astrophysics and Cosmology Research Unit, School of Mathematical Sciences, University of KwaZulu-Natal, Durban 4041, South Africa\\
	}

\date{Accepted XXX. Received YYY; in original form ZZZ}
\pubyear{2018}

\begin{document}
\label{firstpage}
\pagerange{\pageref{firstpage}--\pageref{lastpage}}
\maketitle

%\vspace*{0.5pt}

\begin{abstract}
Satellite galaxies in galaxy clusters represent a significant fraction of the
global galaxy population. Because of the unusual dense environment
of clusters, their evolution is driven by different mechanisms than the ones affecting field or central
galaxies. Understanding the different interactions they are subject
to, and how they are influenced by them, is
therefore an important step towards explaining the global picture of
galaxy evolution.  In this paper, we use the publicly-available high
resolution hydrodynamical simulation Illustris-1 to study
satellite galaxies in the three most massive host haloes (with masses $M_{200} > 10^{14}\,h^{-1}\msun$) at $z=0$. We
measure the Stellar-to-Halo Mass Relation (hereafter SHMR) of the
galaxies, and find that for satellites it is shifted towards lower
halo masses compared to the SHMR of central galaxies. We provide
simple fitting functions for both the central and satellite SHMR. To
explain the shift between the two, we follow the satellite galaxies since their time of accretion into the clusters, and quantify the impact of dark
matter stripping and star formation.  We find that subhaloes start
losing their dark matter as soon as they get closer than $\sim
1.5\times R_{\rm{vir}}$ to the centre of their host, and that up to
80\% of their dark matter content gets stripped during infall.  On the
other hand, star formation quenching appears to be delayed, and
galaxies continue to form stars for a few Gyr after accretion. The combination of these two effects impacts the ratio
of stellar to dark matter mass which varies drastically during infall, from 0.03 to 0.3.

\end{abstract}

%\vspace*{1pt}

\section{Introduction}

Clusters of galaxies are the largest gravitationally bound structures
in the Universe \citep{tormen1998,springel2005,springel2001}.
In the context of the standard hierarchical picture of structure
  formation processes, clusters are the last structures to form
  \citep{lacey93,lacey94,giocoli07}. They reside
  along, and at the nodes of the filamentary network formed by the dark matter density field.
  In these extremely dense regions, galaxies are subject to violent
interactions with their environment, both at the level of dark and
baryonic matter, which forces them to follow particular evolutionary
paths \citep{tormen1998b,tormen2004,gao2004,delucia2004}.

 Many studies show that in the local Universe, galaxies in high density
 environments are mainly red passive ellipticals \citep{oemler1974,
   butcher&oemler1978, dressler1980}, and various mechanisms have been
 identified as having a potential effect on the characteristics of
 galaxies in clusters: ram-pressure stripping \citep{gunn&gott1972}
 can remove the galactic gas and thus quench star formation; frequent
 encounters with other galaxies, called harassment \citep{moore1996,
   moore1998}, can disrupt spiral galaxies into ellipticals; mergers in
 high density environment may favor the survival of massive galaxies
 \citep{merritt85,vandenbosch2005b,conroy07b}; etc. At the same time,
 interactions of dark matter components also drive the evolution
 of infalling galaxies \citep{tormen2004,giocoli2008}.  Numerical
 simulations suggest that dynamical friction sinks galaxies towards
 the center of clusters, with a strength proportional to the mass of
 the galaxy \citep{binney&tremaine2008}. Concurrently, tidal forces of
 the host can strip part of the satellite's matter away, and even
 disrupt it \citep{merritt1983}.

Cosmological simulations offer a privileged tool to follow the
evolution of galaxies in ``real time'' and study the impact of the different interactions they undergo. While
progress in computing speed and development of numerical techniques
allows now to model the evolution of the Universe under the cold
dark matter paradigm, and predicts the structure formation scenario
\citep{springel2005, klypin2011}, the life and evolution of galaxies
remain more demanding to simulate. Indeed, they depend on many complex baryonic
processes acting at different scales. Two main techniques have been
developed in the past decade: semi-analytical
models (hereafter SAMs) or full hydrodynamical simulations.

SAMs \citep{white1991, kauffmann1993,delucia07,somerville2008b,
  guo2010} rely on dark matter simulations, or on Monte Carlo
merger-tree of haloes, populated with seed galaxies. They evolve
following analytical prescriptions motivated by models that sit
between theory and observations.  While this approach has relatively low
computational cost and is quite successful at recovering many
statistical properties of galaxies such as the stellar mass function
\citep{guo2015}, or the gas fraction \citep{somerville2008b}, it does
not directly account for interactions between the baryonic and
dark matter components.  
On the other hand, hydrodynamical simulations
\citep{bonafede11,vogelsberger2014b, schaye2015,deboni18} model the
coevolution of dark and baryonic matter by coupling gravity
with gas physics, and therefore the dynamical processes are more
realistic. They are however much more demanding in terms of
computational power, which strongly limits their volume: the largest
hydrodynamical simulations such as Illustris \citep{vogelsberger2014b}
or EAGLE \citep{schaye2015}, now reach $\sim 100\, \rm{Mpc}$ size,
while dark matter only universes have been simulated in boxes with
side length of up to a few Gpc \citep[eg. the Big MultiDark simulation,
  see][]{klypin2016}.

Here, we want to study the coevolution of dark and baryonic matter.
We use the publicly available Illustris
simulation \footnote{\url{http://www.illustris-project.org}}, one of the state-of-the art hydrodynamical simulations available today. It includes not only gravitational interactions but also gas
dynamics, and some of the most important astrophysical processes, such
as gas cooling, stellar evolution and feedback.  The runs have been
performed with the \textsc{arepo} code \citep{springel10}. The
Illustris simulation was used to study many different aspects of
galaxy evolution, such as their formation \citep{wellons2015,
  martinovic2017}, structure \citep{xu2017} or star formation history
\citep{snyder2015, terrazas2016, bluck2016} among others.  Here we focus on the
evolution of galaxies in clusters, and the evolution of their properties
during accretion processes over cosmic time.

Because of this particular environment, satellite galaxies and
  their subhaloes should evolve differently than central or field
  galaxies, which in the course of their history are continually
  growing through accretion of matter and star formation. Conversely,
   subhaloes will be subject to destructive influence
  from their host, and their dark matter will be gradually stripped
  by tidal forces. This effect has been highlighted in a number of
  analyses of N-body simulations for host haloes ranging from the size
  of the Milky Way \citep{hayashi2003, kravtsov2004, diemand2006,
    buck2019} to that of the most massive galaxy clusters \citep{ghigna1998,
    gao2004, tormen2004, nagai2005, vandenbosch2005b, giocoli2008,
    xie&gao2015, smith2016, rhee2017}, where
    subhalo mass loss is well described by analytical models of tidal stripping
  \citep{mamon2000, gan2010, han2016, hiroshima2018}.

The evolution of the baryonic component has also been widely
  studied. Observations show an increased proportion of red and
  passive galaxies in clusters compared to the field. However, the
  relative importance of the mechanisms that lead to this observation
  are still being debated.
  On one hand, violent interactions such as
  ram-pressure stripping or gravitational interactions, can cause a
  rapid quenching of the satellite galaxies by removing the cold gas
  that fuels the formation of new stars \citep{acreman2003,
    george2013, bahe2015, boselli2016, lotz2018}.  On the other hand,
  some observations favour a slower quenching where the hot gas halo of the galaxies is stripped, preventing their cold gas reservoir  from being replenishing. The cold gas is then consumed gradually, which eventually cause the star formation to stop. 
This slower mechanism is known as 'starvation'
  or 'strangulation' \citep{wolf2009, delucia2012, haines2015,
    zinger2016, tollet2017}.  

 These different evolutionary paths are imprinted notably on
  the Stellar-to-(sub)Halo Mass Relation (SHMR hereafter). As suggested by gravitational lensing measurements \citep{limousin2007, natarajan2009,
    li2015, sifon2015, niemiec2017, sifon2018} or measurements calibrated by
  abundance matching technique \citep{vale&ostriker2004,
    rodriguez2012, rodriguez2013}, the SHMR of satellite galaxies is
  shifted towards lower halo masses compared to that of field galaxies.  With
  the advent of large hydrodynamical simulations that allow to
  self-consistently model the co-evolution of dark and baryonic
  matter, and thus include any baryonic process that
  could affect dark matter evolution, some recent studies have
  re-examined the reason for this shift. \citet{smith2016} found that
  the stellar component of cluster galaxies is affected by stripping
  only when the subhalo has lost a large fraction of its dark
  matter. In their sample, a vast majority of galaxies do not  substantially form
  stars, and only $18\%$ increase their stellar mass by
  more than $15\%$ during infall.  
  Using the same set of simulations,
  \citet{rhee2017} measured a tight relation between time since
  infall, tidal mass loss, and position in the phase-space diagram.
  Finally, \citet{bahe2017} measured the SHMR for galaxies in the
  Hydrangea simulation, and observed a shift for galaxies located as
  far as 5 times the virial radius from the centres of their
  clusters. They argue that this shift is due to the tidal stripping
  of subhaloes for galaxies within $2R_{\rm{vir}}$ from the centre of
  their host, and due to increased star formation for the others.

In this paper, we take advantage of the publicly available
  simulation Illustris, which combines a very high force and mass
  resolution with a relatively large cosmological volume, in order to push further our
  understanding of galaxy evolution in clusters. We measure the
  shift  between the SHMRs of central and satellite galaxies, and
  examine the mechanisms that may cause this difference. This is done for both dark and baryonic matter. We quantify their relative
  importance as well as the time scales over which they operate.  

This paper is organized as follows: Sect.~\ref{sec:data}
presents the data from the Illustris simulation we use;
Sect.~\ref{sec:shmr_ill} details our measurements of the
SHMR for the simulated galaxies (both
centrals and satellites), and presents the best fitting models for the
relation in both cases; Sect.~\ref{sec:tevol} discusses the
evolution of galaxies during infall, making use of the merger trees for
all subhaloes in cluster-like host-halos; Sect.~\ref{sec:m_vs_rsat}
presents the quantification of the stripping of halos as a function of their distance to
the cluster centre; Sect.~\ref{sec:summary} summarizes our results, and discusses them in a wider context.  The cosmology used
throughout this paper is identical to that used in the Illustris
simulation, a flat $\Lambda$CDM universe consistent with the
\emph{Wilkinson Microwave Anisotropy Probe} 9-year data release \citep[WMAP9,
][: $\Omega_{\rm{m,0}}=0.2726$, $\Omega_{\Lambda, 0} = 0.7274$,
  $\Omega_{\rm{b, 0}} = 0.0456$, $\sigma_8 = 0.809, n_{\mathrm{s}} =
  0.963$ and $H_0 = 70.4 \rm{km\,s}^{-1}$]{hinshaw2013}. The notation
log() refers to the base 10 logarithm.

\section{Data}
\label{sec:data}

\subsection{The Illustris simulation}
\label{sec:ill_data}

In this analysis we use the publicly-available data from the
  Illustris simulation, more specifically the group catalogues and
  merger trees.  We summarize in this section the simulation details
  and refer to the corresponding publications.

\paragraph*{Simulation details.} Illustris is a hydrodynamical simulation 
\citep{vogelsberger2014a, vogelsberger2014b} in which dark matter and gas
dynamics evolve simultaneously, using the moving-mesh code
\textsc{arepo} \citep{springel2010}, from initial conditions following
a $\Lambda$CDM cosmology with WMAP-9 parameters \citep{hinshaw2013}
starting at redshift $z = 127$, in a comoving box with a side of
$75\,h^{-1}\mathrm{Mpc} = 106.5\,\mathrm{Mpc}$.  The
simulation includes astrophysical processes to
drive galaxy evolution, including gravity, gas cooling and heating,
stellar formation and evolution, feedback from stars, and supermassive
black holes.
 
Three simulations were run with different resolutions: Illustris-1
(dark matter particle $m_{\mathrm{DM}} = 4.41\times10^6\,h^{-1}\msun$,
baryonic particle $m_{\mathrm{b}} = 8.86\times10^5\,h^{-1}\msun$);
Illustris-2 ($m_{\mathrm{DM}} = 3.53\times10^7\,h^{-1}\msun$,
$m_{\mathrm{b}} = 7.09\times10^6\,h^{-1}\msun$); and Illustris-3
($m_{\mathrm{DM}} = 2.82\times10^8\,h^{-1}\msun$, $m_{\mathrm{b}} =
5.67\times10^7\,h^{-1}\msun$). The work presented in this paper is
based on the highest resolution run, Illustris-1. The simulation is sampled
in 136 snapshots from $z = 127$ to $z=0$. Groups were detected by the Illustris collaboration
with a Friends-of-Friends (FoF) algorithm with linking length $b =
0.2$ , and the haloes were extracted using the \textsc{subfind}
algorithm \citep{springel2001, dolag2009}, and classified into centrals
and satellites from their ranking within their FoF group. Thus, the
central halo is generally the most massive subhalo in the group. The
snapshot at $z = 0$ contains $4, 366, 546$ \textsc{subfind} groups.
 
\paragraph*{Merger trees.} Merger trees were constructed by \citep{rodriguez-gomez2015} for the simulation using two 
different codes, \textsc{SubLink}  and
\textsc{LHaloTree} \citep{springel2005}. 
We chose the \textsc{SubLink} merger
trees for our analysis. They created the trees as follow. At each time
step, a descendant is identified for each subhalo, based on the number
of common particles and their binding energy. To avoid losing low mass
subhaloes when they cross a structure, \textsc{SubLink} allows for one
snapshot to be skipped when it looks for descendants.  Once all the
descendant connections are established the main progenitor of each
subhalo is defined as the one with the most massive history behind it.

\subsection{Cluster haloes and their subhaloes in the Illustris simulation}
\label{sec:ill_clusters}

In this section we describe the haloes and subhaloes from the
Illustris-1 simulation that are used in this work. As described above, the Illustris-1
simulation is the most resolved run, with dark matter particle mass
$m_{\mathrm{DM}} = 6.3\times10^6\,\msun$ and effective baryonic
resolution $m_{\mathrm{b}} = 1.3\times10^6\,\msun$.  We select the three most massive systems in the simulation, with a mass $M_{200} >
10^{14}\,h^{-1}\msun$ at $z = 0$ -- where $M_{200}$ refers to the
  mass enclosing $200$ times the critical density of the universe
  $\rho_c$, with $R_{200}$ the corresponding radius. 
  This mass
selection is equivalent to the redMaPPer cluster
selection with richness $\lambda > 20$ \citep{rykoff2014,
  rozo2014}. We present in Table~\ref{tab:ill_clusters} the main
properties of these haloes at the two redshifts of interest, $z = 0$
and $z = 0.35$, where the latter is the mean redshift of the
redMaPPer-SDSS clusters, used to compare our results to
observations \citep[e.g][]{li2015, niemiec2017}.  In the next section
we compare the SHMR measured for
satellite galaxies of these three host haloes to the one measured for all
central galaxies of the simulation (142,720 centrals at $z=0$ with
$M_* > 0$).

\begin{figure} 
  \begin{center}
    \includegraphics{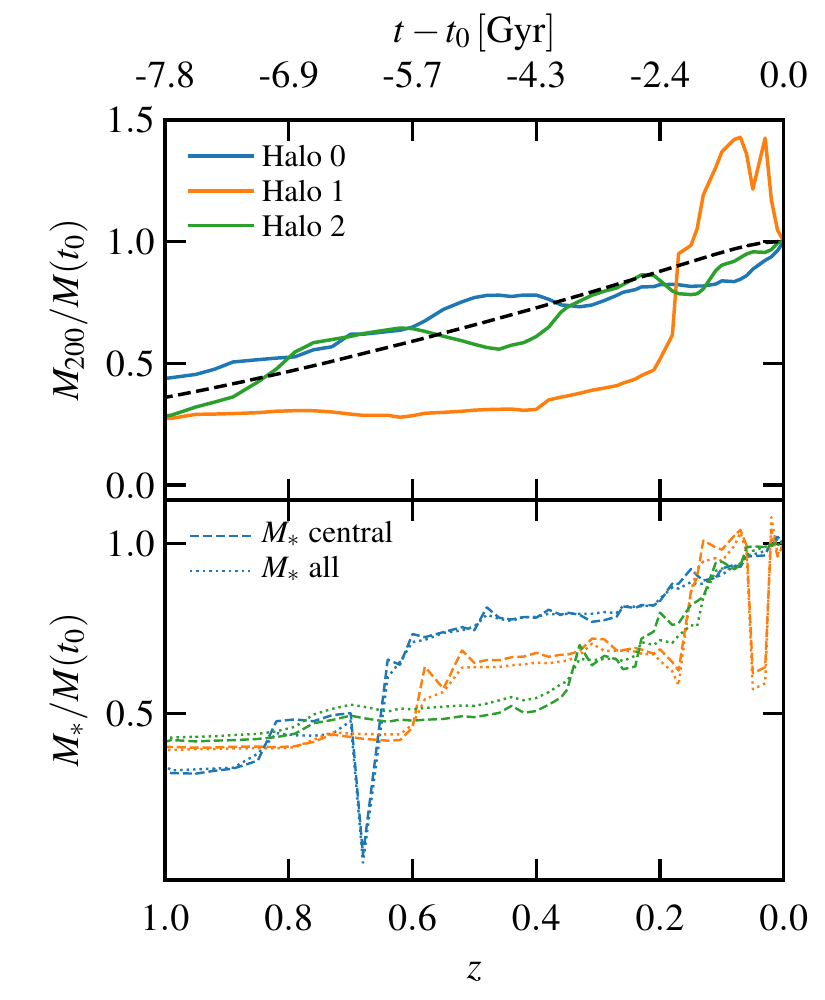}
    \caption{Mass accretion history of the three most massive haloes
      in the Illustris-1 simulation, showing $M_{200}$ in the top panel,
      and the stellar mass in the bottom panel defined as the central
      galaxy stellar mass (dashed line), and the mass of all stellar
      particles in the halo, excluding the subhaloes (dotted
      line). The recent drastic mass changes in the accretion history
      of Halo\,1 suggest that it is undergoing a major merger. We show
      in addition as the black dashed line in the top panel, the
      average mass accretion history for a halo with $M_{200}(z = 0) =
      10^{14} h^{-1}\msun$ as predicted from the model presented by
      \citet{giocoli2012,giocoli2013}. }
    \label{fig:ill_host_evol}
  \end{center}
\end{figure}

The top panel of Fig.~\ref{fig:ill_host_evol} shows the mass evolution
of the three haloes from $z = 1$ to $z = 0$.  One  can
see that Halo\,1 has undergone some violent mass changes in recent
times due to a major merger event.  Looking at the mass history of the
most massive subhaloes in Halo\,1, a recent merger of three haloes of
mass $\sim 1-5 \times 10^{13}\,h^{-1}\msun$ is indeed identified
within the redshift range z=0-0.4.  It is therefore possible that the subbaloes of Halo\,1 experience a different evolution than the subhaloes of Halo\,0 and Halo\,2.  
In the top panel of Fig.~\ref{fig:ill_host_evol}, the dashed line shows the average mass accretion
history predicted from the model presented by
\citet{giocoli2012,giocoli2013}. This relation is consistent with the measured evolution of Halo\,0
and Halo\,2, which have experienced a smoother evolution since $z = 1$ than Halo\,1.

\begin{table*}
\centering
\begin{tabular}{c | c c c c | c c c c}
		& \multicolumn{4}{c}{$z = 0$}													& \multicolumn{4}{c}{$z = 0.35$}											\\
\hline
Halo ID	&	$\log M_{200}/h^{-1}\msun$	&	$\log M_{*}/h^{-1}\msun$	& $N_{\mathrm{subs}}$	&	$N_{\mathrm{sub}}^*$	&	$\log M_{200}/h^{-1}\msun$	& $\log M_{*}/h^{-1}\msun$	& $N_{\mathrm{subs}}$	&	$N_{\mathrm{subs}}^{*}$	\\
\hline
0		&	14.21		&	12.18		& 	5070				&	120				&	14.08 		& 12.08		& 4113				&	97				\\
1		&	14.20		&	11.99		&	6756				&	138				&	13.76 		& 11.81		& 1855				&	33				\\
2		&	14.19		&	12.13 		&	5262				&	112				&	14.05 		& 11.89		& 4268				&	85				\\
		
\hline
\end{tabular}
\caption{Properties of the three most massive haloes in the Illustris
  simulation: mass $M_{200}$, stellar mass of the central galaxy
  $M_{*}$ defined as the sum of all star particles within twice the
  stellar half mass radius, the total number of subhaloes
  $N_{\mathrm{sub}}$, and the number of subhaloes with $M_{\mathrm{sub}} >
  10^{10}\,h^{-1}\msun$ $N_{\mathrm{sub}}^*$.}
\label{tab:ill_clusters}
\end{table*}

\begin{figure} 
  \begin{center}
    \includegraphics{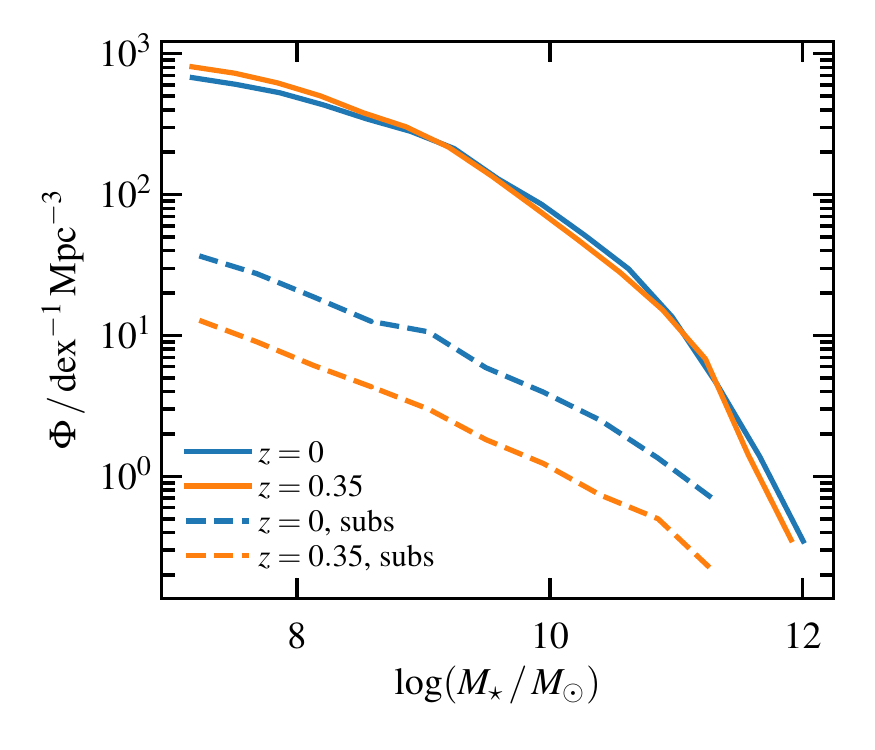}
    \caption{Stellar mass function for all the central galaxies in the
      Illustris-1 simulation (solid line), and the satellite galaxies
      of the three most massive haloes (dashed line), at $z = 0$
      (blue) and $z = 0.35$ (orange.)  }
    \label{fig:SMF}
  \end{center}
\end{figure}

Different estimates of the mass are available for the various objects
in the simulation. Following \citet{vogelsberger2014b}, we use galaxy
properties defined within twice the stellar half-mass radius. The
stellar mass is therefore the sum of the mass of all star particles
within this radius. We define stellar mass in the central haloes as the
stellar mass in the central galaxy (another definition could be the
sum of all star particle in the halo excluding the subhaloes; we show
the stellar mass accretion history for these two definitions in the bottom
panel of Fig.~\ref{fig:ill_host_evol}). Fig.~\ref{fig:SMF} shows
the Stellar Mass Function (SMF) of all the central galaxies in the
simulation (solid line), and of the satellite galaxies of the three host
haloes described above (dashed line), at $z = 0$ (blue) and $z = 0.35$
(orange).  One can see that while the satellite SMF
increases by $\sim 0.5$\,dex between $z = 0.35$ and $z=0$, the central
SMF does not evolve significantly.

We remind the reader that for the dark matter content, central haloes
are defined as spherical regions with a radius $R_{200}$, with an
average density equal to 200 times the critical density of the
Universe, $\rho_{\mathrm{c}}(z)$.  The mass of the central halo is the
total mass enclosed in this region, $M_{200}$. For the subhaloes, where this definition does not apply,
the mass is defined as the sum of the masses of all particles
identified as being gravitationally bound to the subhalo
\citep{springel2001,gao2004,giocoli2008,giocoli2010}.

\section{Stellar-to halo mass relation of haloes and subhaloes}
\label{sec:shmr_ill}

\subsection{SHMR for central haloes}
\begin{figure*} 
  \begin{center}
    \includegraphics{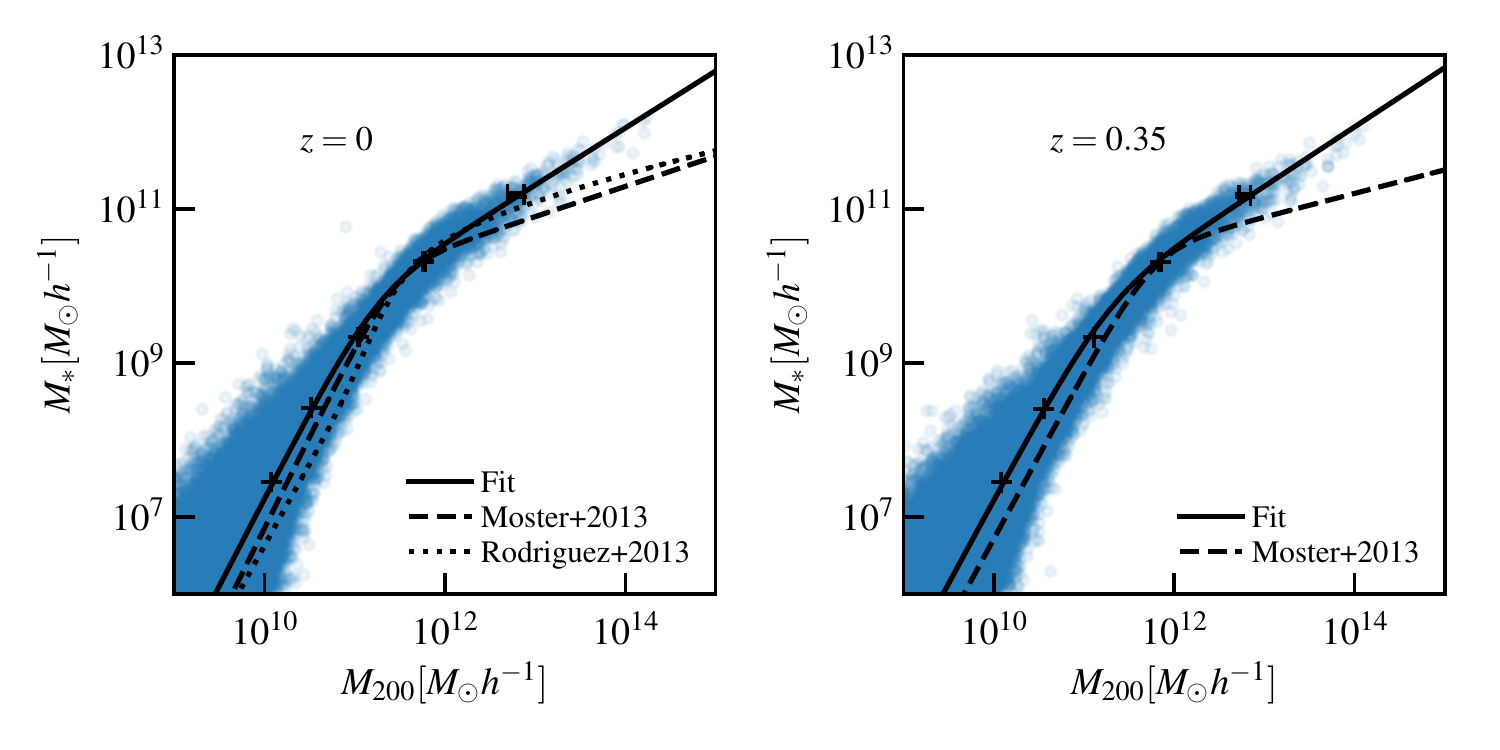}
    \caption{Stellar-to-halo mass relation for haloes from the
      Illustris-1 simulation.  The left and right panels show the
      measurements at redshift $z=0$ and
      $z=0.35$, respectively. The masses are expressed in units of $h^{-1}\msun$. Blue
      open circles represent the SHMR for each central halo of the
      simulation. The black points are median values in bins of
      stellar mass. The error bars represent the $1\sigma$ uncertainty on the median. In addition, we plot the relation computed from
      abundance matching in \citet{moster2013} at the two redshifts
      (dashed line), and in \citet{rodriguez2013} for $z=0$ only (dotted
      line). The black solid lines represent the best-fit relation. }
    \label{fig:shmr_halo_ill}
  \end{center}
\end{figure*}

We now focus on the stellar-to-halo mass relation (SHMR). 
We first look at the relation for central haloes. This will be used as our reference for the comparison with subhaloes.
The relation is shown in Fig.~\ref{fig:shmr_halo_ill}.
The blue open circles mark the SHMR for each (central) halo of the
simulation, and in black points we plot the median relation in five bins
of stellar mass: $10^7 < M_* < 10^8$, $10^8 < M_* < 10^9$, $10^9 < M_*
< 10^{10}$, $10^{10} < M_* < 10^{11}$ and $10^{11} < M_* < 10^{12}$,
in units of $h^{-1}\msun$, where the small error bars represent
the 1$\sigma$ uncertainty on the median.
Comparing the two panels, we do not notice a strong evolution between
redshift $z=0$ and $z = 0.35$, in agreement with the SMF described
above, which shows that stars and dark matter evolve together.

In addition, we show the stellar-to-halo mass relation obtained from
abundance matching in \citet{moster2013}, defined as:
\begin{equation}
\frac{M_{*}}{M_{200}} = 2N\left[
  \left(\frac{M_{200}}{M_1}\right)^{-\beta} +
  \left(\frac{M_{200}}{M_1}\right)^{\gamma}\right]^{-1}
\label{equ:moster2013}
\end{equation},
where the best fit parameters from \citet{moster2013} at $z=0$ are
$\log(M_1/\msun) = (11.590 \pm 0.236)$, $N = (0.0351 \pm 0.0058)$,
$\beta = (1.376 \pm 0.153)$ and $\gamma = (0.068 \pm 0.059)$
\citep[the redshift dependence of the parameters is given in
][]{moster2013}. We also plot the relation obtained with abundance
matching in \citet{rodriguez2013} for central galaxies (we use their
results from set C).

These two SHMR differ from our measurements at the two mass extrema.
This discrepancy is due to shortcomings in both calculations. It has been
shown \citep{sawala2013, munshi2013} that SHMR estimated from dark
matter only simulations \citep[as in][]{moster2013} overestimates the
mass of dark matter haloes, especially at low masses, due to the lack
of baryonic physics. However, when correcting for this difference by
matching the subhaloes to their counterpart in Illustris DM-only run,
the Illustris simulation still appears to overpredict the stellar
masses of the most and least massive haloes \citep{genel2014,
  vogelsberger2014b}. At the high mass end, the prediction of overly
massive galaxies could be due to insufficient AGN feedback in the
Illustris samples \citep[eg.][]{vogelsberger2014b}.  At
intermediate mass scales, $M_{*} \sim 10^9 - 10^{11}\,h^{-1}\msun$,
the measurements are in good agreement.

We fit the parameters $M_1$, $N$, $\beta$ and $\gamma$ from
equation~\ref{equ:moster2013} to the measured SHMR at $z=0$ and $z=0.35$. As our
galaxy sample is dominated by low-mass objects, we need to weight
their contribution. We therefore split the galaxies into 15 bins in
$\log M_*$, and perform fit the median values in each bin, with
the standard deviation of $\log M_*$ in each bin as the error
estimate. We obtain the best-fit parameters and the intervals of
confidence with a Markov Chain Monte Carlo (MCMC) method using
\textsc{emcee} \citep{foreman-mackey2013}, which is a Python
implementation of an affine invariant MCMC ensemble sampler.  The
best-fit parameters and the 68\% credible intervals are presented in
Table~\ref{tab:shmr_fits}. We also present the joint 2-dimensional, and marginalized
1-dimensional posterior probability distributions for the parameters
$M_1$, $N$, $\beta$ and $\gamma$ at $z = 0$ in
Appendix~\ref{sec:corner_plots}. The ``normalization'' parameters
$M_1$, and $N$, are in reasonable agreement with \citet{moster2013}, but
as expected the slopes are different, steeper at high mass and flatter
at low mass. The solid black curve shows the SHMR corresponding to our
best-fit parameters in Fig.~\ref{fig:shmr_halo_ill}.  
To compare our measurements with \citet{moster2013}, we let our best-fit SHMR parameters evolve bewtween $z=0$ and $z = 0.35$ using the redshift parametrization described in equations (11) - (14), and Table 1 from
\citet{moster2013}. The values are shown in Table~\ref{tab:shmr_fits}. We measure a weaker redshift evolution (consistent with no evolution) than what \citet{moster2013} predicts.

\begin{table}
\centering
\begin{tabular}{|c c c c|}
\hline
\multicolumn{4}{| c |}{Haloes}  \\
\hline
                        & Fit $z = 0$               &  Fit $z = 0.35$           & Evolved   \\
\hline
$\log M_1/h^{-1}\msun$  & $11.21^{+0.18}_{-0.17}$   & $11.26^{+0.20}_{-0.20}$   & $11.52$   \\
$N$                     & $0.030^{+0.007}_{-0.006}$ & $0.025^{+0.006}_{-0.006}$ & $0.024$   \\
$\beta$					& $1.27^{+0.13}_{-0.11}$	& $1.17^{+0.13}_{-0.11}$    & $1.06$    \\
$\gamma$				& $0.26^{+0.08}_{-0.08}$	& $0.23^{+0.09}_{-0.08}$    & $0.35$    \\
\hline\hline
\multicolumn{4}{| c |}{Subhaloes}   \\
\hline
                        & Fit $z = 0$               &  Fit $z = 0.35$           & Evolved   \\
\hline
$\log M_1/h^{-1}\msun$  & $10.66^{+0.35}_{-0.26}$   & $10.67^{+0.31}_{-0.24}$   & $10.97$   \\
$N$                     & $0.091^{+0.034}_{-0.021}$ & $0.099^{+0.033}_{-0.020}$ & $0.084$   \\
$\beta$					& $0.98^{+0.13}_{-0.12}$    & $0.95^{+0.11}_{-0.11}$    & $0.78$    \\
$\gamma$				& $0.16^{+0.17}_{-0.09}$    & $0.09^{+0.14}_{-0.06}$    & $0.25$    \\
\hline
\end{tabular}
\caption{ Best-fit parameters and $68\%$ credible intervals for
  the parameters $M_1$, $N$, $\beta$ and $\gamma$ from
  equation~\ref{equ:moster2013} fitted to the haloes and subhaloes at
  $z=0$ and $z = 0.35$. We assumed flat priors such as: $\log M_1/h^{-1}\msun \in [10.5;12.5]$, $N \in [0;2]$, $\beta \in [0;5]$ and $\gamma \in [0;5]$. We also present the value of the $z=0$ best fit parameters evolved to $z = 0.35$ using the redshift parametrization from \citet{moster2013}. }
\label{tab:shmr_fits}
\end{table}

\subsection{SHMR for subhaloes}
\begin{figure*} 
  \begin{center}
    \includegraphics{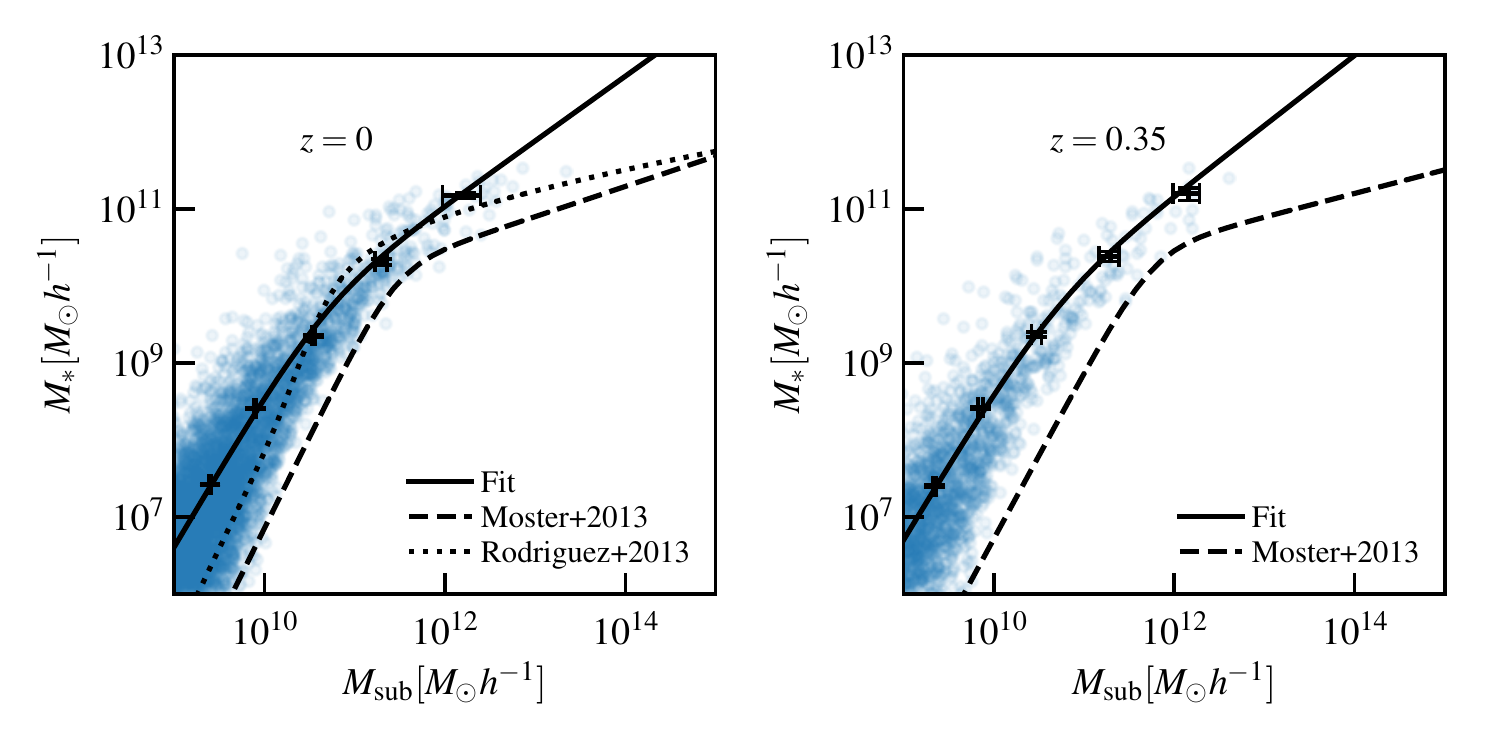}
    \caption{Stellar-to-halo mass relation for subhaloes of the
      ``cluster like'' haloes from the Illustris-1 simulation. Left and right panels shows our measurements at $z=0$, and 
      $z=0.35$ respectively. In addition, we plot our best-fit
      relations at $z = 0$ and $z = 0.35$ (solid line). For comparison, we show the same relation as in Fig.~\ref{fig:shmr_halo_ill}, computed from simulations
      in \citet{moster2013} (dashed line). The relation for subhaloes
      appears to be shifted towards lower dark matter mass compared to
      the one for haloes. We also display the relation from
      \citet{rodriguez2013} at $z=0$, measured by abundance matching
      for satellite galaxies and their subhaloes (dotted line). }
    \label{fig:shmr_subhalo_ill}
  \end{center}
\end{figure*}

We now plot the SHMR for subhaloes in the three cluster-like haloes
described in Sect.~\ref{sec:ill_clusters}. 
We consider all subhaloes in the \textsc{subfind} catalogues, except for the first one as it is the central halo.
Figure~\ref{fig:shmr_subhalo_ill} shows the relation for individual subhaloes,
in stellar mass bins, similarly to the SHMR for central halos.

Figure~\ref{fig:shmr_subhalo_ill} shows that the SHMR is shifted towards lower halo
masses for subhaloes compared to central haloes. In
Table~\ref{tab:shmr_ill}, we list the stellar and halo masses for each bin, for both
central and satellite populations. The error bars represent the
standard errors. We fit again the parameters $M_1$, $N$, $\beta$ and
$\gamma$ from equation~\ref{equ:moster2013} using the same procedure as for centrals. The best fit values and the $68\%$
credible intervals at $z =0$ and $z = 0.35$ are listed in Table~\ref{tab:shmr_fits}. The posterior
probability distributions at $z = 0$ are given in Appendix~\ref{sec:corner_plots}.  The
solid line in Fig.~\ref{fig:shmr_subhalo_ill} shows the relation
constructed using the best fit parameters. In Table~\ref{tab:shmr_fits}, we also list the best fit parameters at $z=0$ evolved to $z=0.35$, adopting the \citet{moster2013} redshift parametrization. As in the case of central galaxies, our measurements show a weaker redshift evolution than \citet{moster2013}.

The dotted line in Fig.~\ref{fig:shmr_subhalo_ill} shows the SHMR for satellite galaxies measured by
abundance matching in \citet{rodriguez2013}. The shift of this
relation compared to central galaxies is similar to our measurement at
intermediate mass scale ($M_* \sim 10^9 - 10^{11}\,h^{-1}\msun$), but
the slopes differ at both mass ends.

\begin{figure} 
  \begin{center}
    \includegraphics{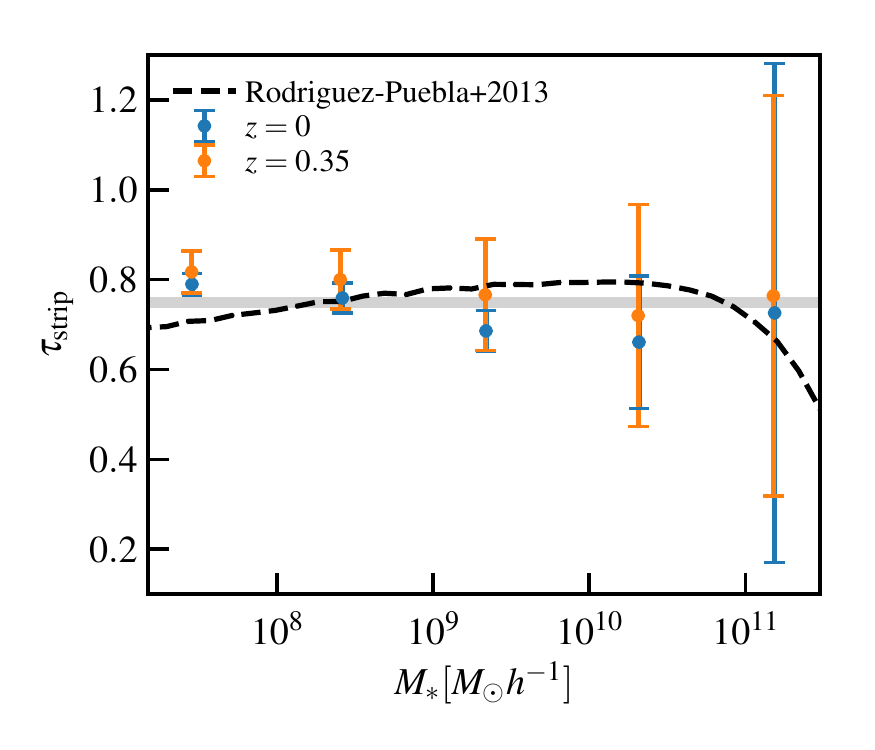}
    \caption{Stripping factor $\tau_{\mathrm{strip}}$ as a function of
      the mean stellar mass in each bin at $z = 0$ (blue), and at $z = 0.35$ (orange). The grey line represents the mean value,
      and the dashed line is the stripping factor computed using the
      SHMR from \citet{rodriguez2013}.}
    \label{fig:tau_evol_ill}
  \end{center}
\end{figure}

Finally, considering the assumption that the stellar mass does not evolve
during accretion, and that the ``progenitors'' of subhaloes at a given stellar mass are central haloes of same stellar mass, we
define the stripping factor as
\begin{equation}
\tau_{\mathrm{strip}}(M_*) = 1 - \frac{M_{\mathrm{sub}}(M_*)}{M_{\mathrm{h}}(M_*)}\mathrm{,}
\label{equ:tau_strip}
\end{equation}
and present the results for each stellar mass bin in
Table~\ref{tab:shmr_ill}. From this perspective, the stripping factor
simply represents the shift in halo mass of the SHMR between central
haloes and subhaloes. Fig.~\ref{fig:tau_evol_ill} shows the evolution of the stripping factor as
a function of the mean stellar mass in each bin for $z = 0$ and $z =
0.35$. There is no significant
evolution with the stellar mass nor the redshift. In addition, we plot
the mean value of the stripping factor $\langle \tau_{\rm strip}
\rangle = 0.75$, and find no significant deviation from it. We
  compare our results with those obtained by \citet{rodriguez2013}
  using abundance-matching by computing $\tau_{\rm{strip}}$ from their
  SHMR for satellite and central galaxies. Although they find a mass
  dependence for the stripping factor, it is small in the mass range
  that we consider, and we consider our results to be in good agreement. We also note that the relation from \citet{rodriguez2013} was calibrated using dark matter-only simulations, where the impact of baryons on halo formation history is not taken into account. This could explain the small difference that we observe at the low mass end.   
  Finally we checked that defining the mass of the central haloes as the sum of the masses of all gravitationally bound particles $M_{\mathrm{bound}}$, in order to use the same dark matter mass definition as for subhaloes, or using $M_{*\mathrm{,bound}}$ for both haloes and subhaloes, does not change our conclusions.

\begin{table}
\centering
\begin{tabular}{c | c c c  }
				& \multicolumn{3}{c}{$z = 0$}																					\\
\hline
$M_*$ bin			&	$M_{200}\times 10^{-11}$	&	$M_{\mathrm{sub}} \times 10^{-11}$	&	$\tau_{\mathrm{strip}}$	\\
($h^{-1}\msun$)		&  ($h^{-1}\msun$)				&	($h^{-1}\msun$)									 &							 \\
\hline
$10^7 - 10^8$		&	$0.119 \pm 0.001$			&	$0.025 \pm 0.001$					&	$0.79 \pm 0.02$			\\
$10^8 - 10^9$		&	$0.331 \pm 0.001$			&	$0.080 \pm 0.003$					&	$0.76 \pm 0.03$			\\
$10^9 - 10^{10}$	&	$1.10 \pm 0.01$				&	$0.347 \pm 0.020$					&	$0.69 \pm 0.05$			\\
$10^{10} - 10^{11}$	&	$5.83 \pm 0.21$				&	$1.98 \pm 0.37$						&	$0.66 \pm 0.15$			\\
$10^{11} - 10^{12}$	&	$62.4 \pm 13$				&	$17.1 \pm 9.6$						&	$0.73 \pm 0.56$			\\
\hline
\hline
				& \multicolumn{3}{c}{$z = 0.35$}																				\\
\hline
$10^7 - 10^8$		&	$0.121 \pm 0.001$			&	$0.022 \pm 0.001$					&	$0.82 \pm 0.05$		\\
$10^8 - 10^9$		&	$0.358 \pm 0.001$			&	$0.071 \pm 0.006$					&	$0.80 \pm 0.06$		\\
$10^9 - 10^{10}$	&	$1.28 \pm 0.01$				&	$0.300 \pm 0.046$					&	$0.77 \pm 0.12$		\\
$10^{10} - 10^{11}$	&	$6.99 \pm 0.24$				&	$1.96 \pm 0.60$						&	$0.72 \pm 0.25$		\\
$10^{11} - 10^{12}$	&	$61.0 \pm 11.0$				&	$14.4 \pm 5.8$						&	$0.76 \pm 0.44$		\\

\end{tabular}
\caption{Mean masses of central haloes and subhaloes for each
  stellar mass bin, at $z = 0$ and $z = 0.35$. The stripping factor in
  each bin is also given.}
\label{tab:shmr_ill}
\end{table}

\section{Evolution since the time of accretion}
\label{sec:tevol}

In this section, we investigate which mechanisms drive this shift of
the SHMR towards lower halo masses for subhaloes. Is it an evolution of
the dark matter mass at fixed stellar mass, completely dominated by
tidal stripping?  Or is there a contribution from ongoing star
formation?  How does the tidal stripping and star formation quenching
timescales compare?  Is there a significant contribution from mergers?  
To start answering these questions we will follow the evolution of
subhalo properties from the time of accretion to present time.

We choose again the three most massive haloes of 
Illustris-1 (with $M_{\rm h} > 10^{14}\,h^{-1}\msun$) as cluster-like
haloes, and we examine the evolution of their subhaloes. We only select subhaloes with $M_{\mathrm{sub}} > 10^{10}\,h^{-1}\msun$, to
guaranty a sufficient number of particles (i.e $N_{\rm{part}} >
280$) to ensure that the subhaloes are reasonably resolved above the mass and force resolution of the simulation.
We follow the evolution of the subhaloes with time, by
extracting the main branch of their merger trees. We use the merger
trees obtained with the \textsc{SubLink} algorithm as described in
Sect.~\ref{sec:ill_data}.

We need to define a time of reference, when the subhalo
starts its accretion into the host halo.
This accretion time $t_{\mathrm{acc}}$ is defined as the \textit{first} time the halo enters the shell of radius
$R_{\mathrm {acc}}$.  We define the accretion radius as twice the virial
radius of the host halo, $R_{\mathrm {acc}} = 2\times R_{200}$. Indeed,
the cluster environment extends far beyond the virial
radius, and its influence on the infalling subhaloes can therefore
start before they reach $R_{200}$. A more physically motivated choice
for the accretion radius would be to use the splashback radius
\citep{more2015,bush2017,baxter2017,diemer2017}, which is estimated by
current measurements at around 1.5-2$\times R_{200}$, which motivates
our choice of $R_{\rm{acc}} = 2\times R_{200}$.

The subhaloes are then followed snapshot by snapshot after $t_{\mathrm{acc}}$.
In addition, as the measurement of the subhalo mass across the snapshots
can be noisy \citep{muldrew2011}, we perform sigma-clipping in order to clean
the mass evolution signal.
	
\subsection{Evolution of the halo-centric distance} 
\begin{figure*} 
  \begin{center}
    \includegraphics{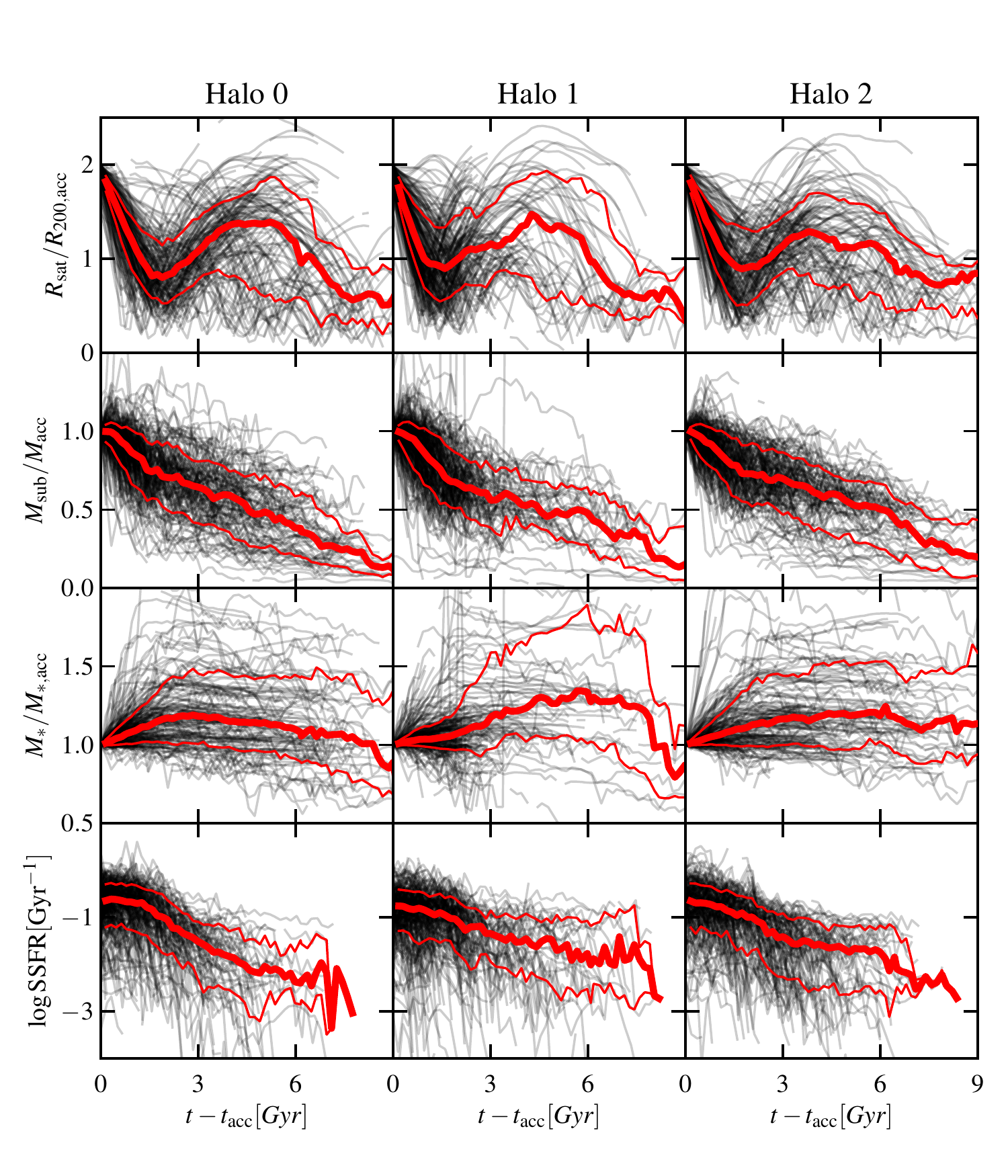}
    \caption{Evolution of subhalo properties as a function of time
      since accretion: halo-centric distance normalized by the host
      virial radius at accretion (\textit{top panel}), subhalo dark
      matter mass normalized by the dark matter mass at accretion
      (\textit{second panel}), stellar mass normalized by stellar mass
      at accretion (\textit{third panel}), and specific star formation
      rate (SSFR) (\textit{bottom panel}). Each column represents one
      of the cluster-like haloes. The black lines represent the
      evolution of each subhalo independently; the thick red
      line shows the median evolution, and the thin red lines mark the 16th
      and 84th percentiles.}
    \label{fig:all_evol}
  \end{center}
\end{figure*}

The first characteristic of interest is the
distance between the subhalo and the centre of the host halo. The top panel of Fig.~\ref{fig:all_evol} shows the evolution of
the cluster-centric distance normalized by the virial radius of the
host at the time of accretion, as a function of time since accretion, for each of the three cluster-like haloes
separately (each column corresponds to a different host). In each panel, the
black curves represent the evolution for each subhalo independently,
the thick red line indicates the median value at
each time step, and the thin red lines highlight the evolution of the
16th and 84th percentiles.

As expected, the subhaloes are moving towards the centre of their
hosts, and this overall infall motion has already started at $2\times
R_{200}$ from the centre. It is therefore interesting to follow the
properties of the subhaloes at this distance, as the majority of them
will be accreted, and will end up in the central part of the cluster
\citep{nipoti18}. One will note that some of them are already influenced
by the cluster at such distance. In general, subhaloes are following
spiraling orbits typically off-centered, leading them towards the
centre of the cluster \citep{hayashi03,hayashi07}.

We can also observe the first infall into the cluster, where the
subhaloes pass close to the centre of the cluster. We measure for each
host halo the time at which the average $R_{\rm{sat}}/R_{200,
  \rm{acc}}$ evolution reaches its first minimum:
$t_{\rm{min}} = 1.8\,\rm{Gyr}$ for Halo\,0, $t_{\rm{min}} =
1.5\,\rm{Gyr}$ for Halo\,1, and $t_{\rm{min}} = 1.7\,\rm{Gyr}$ for Halo\,2.  This
first crossing has been discussed in some studies 
\citep[see for example][]{jaffe2015} as the moment when the subhaloes are ram-pressure
stripped from their gas. We will compare this timescale with the dark
matter loss and star formation quenching timescales in the next
sections.

\subsection{Evolution of the dark matter mass} 
We now follow the evolution of the subhaloes dark matter content
during infall. The top middle panel of Fig.~\ref{fig:all_evol} shows
the time evolution of the mass of dark matter particles bound to
the subhaloes normalized by the mass at accretion,
$M_{\mathrm{sub}}/M_{\mathrm{acc}}$, as a function of time. We show the evolution for each subhalo and the median value at each time
step.

Looking at the median evolution, the mass normalized by the mass at
accretion shows a decrease over time, starting very soon after the
subhaloes infall, at $2\times R_{200}$. We investigate that case more carefully, and discuss the mass-loss start time in
Sect.~\ref{sec:m_vs_rsat}.  One can see two possible different regimes in
the mass-loss, with the subhaloes losing matter more rapidly up to $t
- t_{\mathrm{acc}} \sim 1.5\,\rm{Gyr}$. To check this trend, we fit a
broken line to the mass evolution, to measure the two slopes of the
evolution, $\alpha_{\rm{dm}}$ and $\beta_{\rm{dm}}$, as well as the time at
which the slope changes, $t_{\rm{dm}}$.  The function is defined as:
\begin{equation}
	\frac{M_{\rm{sub}}}{M_{\rm{acc}}}(t) = \begin{cases} 
    	\alpha_{\rm{dm}}t + c_{\rm{dm}}, & \mbox{if } t < t_{\rm{dm}} \\ 
    	\beta_{\rm{dm}}t + c'_{\rm{dm}}, & \mbox{if } t > t_{\rm{dm}} \end{cases}
        \label{equ:broken_line}
\end{equation},
where $c' = (\alpha_{\rm{dm}} - \beta_{\rm{dm}})t_{\rm{dm}} +
c_{\rm{dm}}$.  This is shown in the top panel of
Fig.~\ref{fig:all_fit}. We perform the fit on the median evolution
over the subhaloes from the three host haloes. The best fit parameters
are presented in Table~\ref{tab:fit_evol}.

The best fit evolution shows a slope change at $t_{\rm{DM}} = (1.72
\pm 0.04)\,\rm{Gyr}$: on average subhaloes appear to lose their mass
faster during their first infall, with a rate of $\sim 20\%$ of their
mass at accretion per Gyr. The mass loss then slows down to a rate of $\sim
6\%$ of their mass at accretion per Gyr.  As shown in previous studies
\citep[eg.][]{diemand2006}, the subhaloes lose most of their mass at
their successive passages at the pericenter, with a relatively larger
fraction at the first passage.

As a comparison we compute a simple analytical model for the subhalo
mass loss by tidal stripping. At each time step, the total mass that is
gravitationally bound to the subhalo can be defined as being enclosed
in the tidal radius, $r_{\rm t}$. Beyond this radius, matter is
disrupted by the tidal forces of the host halo:
\begin{equation}
F_{\rm T} = -\frac{\rm d}{\mathrm{d}R_{\rm sat}} \left( \frac{GM_{\rm{host}}(R_{\rm{sat}})}{R_{\rm{sat}}^2}\right)r_{\rm t}\ , 
\end{equation}
as they are stronger than the subhalo self-gravity:
\begin{equation}
F_{\rm G} = \frac{GM_{\rm{sub}}(r_{\rm t})}{r_{\rm t}^2}\ ,
\end{equation},
where $G$ is the gravitational constant.  For each subhalo, and at each
time step, we compute the value of the tidal radius by solving $F_{\rm
  T} = F_{\rm G}$, assuming that the (sub)haloes follow a
Navarro-Frenk-White density profile \citep[NFW, ][]{nfw1996}, to
compute the mass. The subhalo mass is then defined as the NFW mass
truncated at $r_{\rm t}$. The top panel of Fig.~\ref{fig:all_fit}
shows as a dotted line the evolution of this theoretical value
normalized by mass at accretion, and averaged over all the subhaloes of the
three hosts.

The mass loss predicted by this simple analytical model is in very
good agreement with the simulation during the first infall, i.e up to
$\sim 2-3\,\rm{Gyr}$. However, after that it underestimates the mass
loss. This is to be expected with such a simple model, as for instance
it does not take into account the possible reorganization of the mass
within the subhalo but considers that it keeps following a NFW mass
distribution with an abrupt truncation at the tidal radius. 
Collisions between satellites might also play a role in the mass
evolution, and are not included in such a simple model
\citep{tormen1998b}.

In the top panel of Fig.~\ref{fig:all_fit}, we also show the
exponential mass loss from eq.~(10) in \citet{giocoli2008}. \citet{giocoli2008} adopt a
different definition for the accretion time, namely the time when the
subhalo crosses the virial radius of the host for the last time,
which corresponds in our case to $t - t_{\rm{acc}} \sim 6\, \rm{Gyr}$
(see top panel of Fig.~\ref{fig:all_evol}). We also let the parameter
$\tau_0$ from their equation to vary, as subhaloes survive longer in
simulations that include baryons. It appears indeed that the model
from \citet{giocoli2008} describes quite well the evolution that we
measure after $t - t_{\rm{acc}} = 6\, \rm{Gyr}$, when we fix $\tau_0 =
2.5 - 3 \, \rm{Gyr}$.

Figure~\ref{fig:all_evol} shows that subhaloes 
that remain the longest in the host, up to $t = t_{\mathrm{acc}} +
9\,\rm{Gyr}$, appear to lose $\sim 75\%$ of their mass. However, most of the subhaloes
at $t=0$ have not started their infall into the cluster so long ago, and have therefore lost
less than $75\%$ of their mass. To quantify this effect, we split the
subhaloes into four samples according to their surviving mass at
$z=0$, $f_{\rm{surv}} = M_{\rm{sub}}(z =
0)/M_{\rm{sub}}(z_{\rm{acc}})$, and compute for each sample the mean
time since accretion. As expected, we find that subhaloes with the
lowest surviving mass ($f_{\rm{surv}} < 0.25$) have spent more
time in their host ($t_0 - t_{\rm{acc}} \sim 8\, \rm{Gyr}$, where $t_0$ represents the present time), while
subhaloes with a high surviving mass ($f_{\rm{surv}} > 0.75$) have
started their infall much more recently ($t_0 - t_{\rm{acc}} \sim 2\,
\rm{Gyr}$). Table \ref{tab:tacc_fsurv} summarizes the values obtained for each sample considered.

\begin{table}
\centering
\begin{tabular}{|c  | c c | }
\hline
$f_{\rm{surv}}$	& $\langle f_{\rm{surv}} \rangle$	& $\langle t_0 - t_{\rm{acc}} \rangle$ (Gyr)	\\
\hline
$[0, 0.25]$		& $0.15 \pm 0.07$					& $7.72 \pm 2.11$						\\
$[0.25, 0.5] $	& $0.40 \pm 0.07$					& $5.41 \pm 2.10$						\\
$[0.5, 0.75]$	& $0.62 \pm 0.07$					& $3.71 \pm 1.80$						\\
$[0.75,\infty]$	& $0.94 \pm 0.21$					& $1.92 \pm 1.41$						\\
\hline
\end{tabular}
\caption{Mean surviving mass, $f_{\rm{surv}}= M_{\rm{sub}}(z =
  0)/M_{\rm{sub}}(z_{\rm{acc}})$ at redshift $z = 0$, and mean time
  since accretion, $t_0 - t_{\rm{acc}}$, for subhaloes divided in bins
  of surviving mass. The error bars represent the standard
  deviation. }
\label{tab:tacc_fsurv}
\end{table}

\subsection{Stellar mass evolution} 
We now investigate whether the evolution of the stellar mass during infall could be partly responsible for the SHMR shift.
Similar to the dark matter mass, we normalize the stellar
mass at each epoch by the stellar mass at accretion; we present in
the third panel of Fig.~\ref{fig:all_evol} the normalized stellar mass
as a function of time for the three host haloes. The median stellar
mass increases after the subhaloes cross $2\times R_{200}$,
showing that on average galaxies still have ongoing star forming.
	
\begin{figure} 
  \begin{center}
    \includegraphics{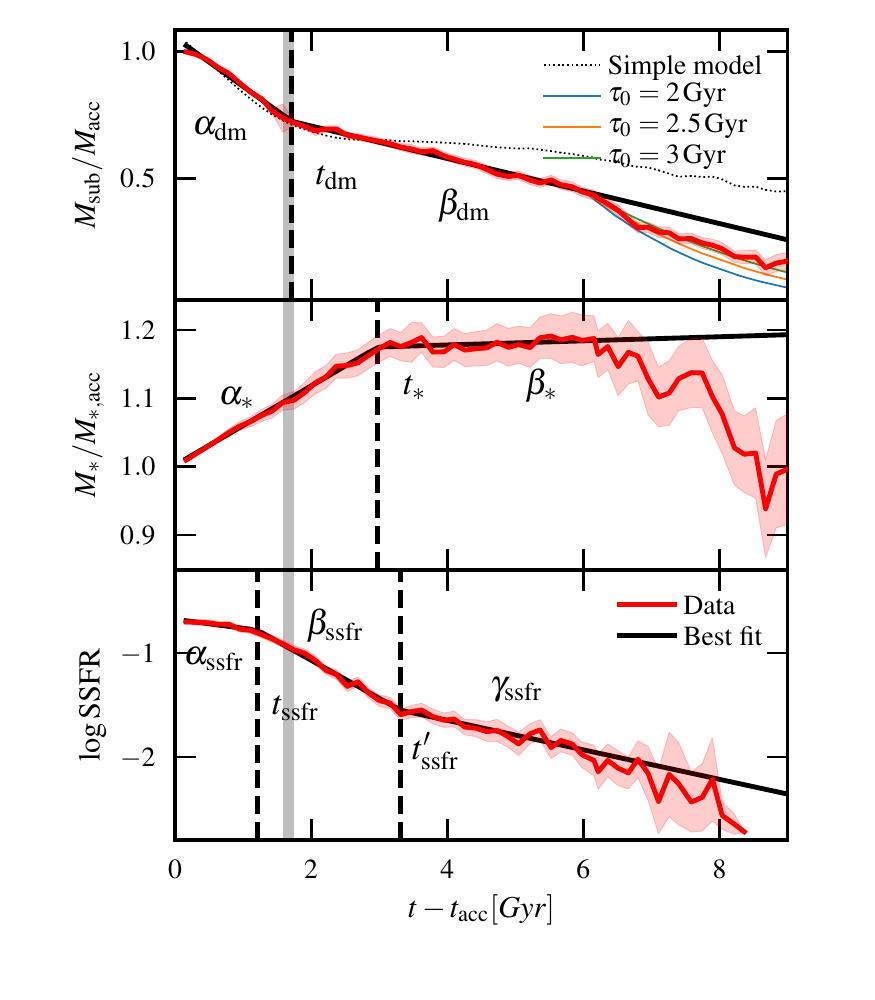}
    \caption{Median evolution of $M_{\mathrm{sub}}/M_{\mathrm{acc}}$
      (top panel), $M_{\mathrm{*}}/M_{\mathrm{*, acc}}$ (middle panel),
      and SSFR (bottom panel), as a function of time (red line, with
      the error on the median as red filled surfaces), and the best fit
      evolution (black solid lines). The grey vertical line
      represents the time at which the mean evolution of the
      cluster-centric distance reaches its first minimum. The dotted
      line in the top panel represents the mass evolution from the dark
      matter stripping simple analytical model (see text for details),
      and the solid blue, orange and green lines correspond to the
      subhalo mass loss measured in \citet{giocoli2008} with $\tau_0=2$, 2.5, and 3 Gyr respectively.}
		\label{fig:all_fit}
  \end{center}
\end{figure}

The median stellar mass then starts to stagnate, demonstrating that the star
formation is slowing down after accretion. Similarly to the dark
matter investigation, we measure the mean evolution over the three host haloes,
and fit a broken line to it, defined as in equation
\ref{equ:broken_line}.  The median evolution is presented in the
middle panel of Fig.~\ref{fig:all_fit}, and the best fit parameters
are listed in Table~\ref{tab:fit_evol}. The fit  shows a
transition between a regime where the stellar mass increases ($+6\%$
of the infall mass per Gyr), and a stagnation ($+0.3\%$ of the infall
mass per Gyr) at $t_* = (2.98 \pm 0.06)\,\rm{Gyr}$.

To see the transition more clearly, we show in the bottom panel of
Fig.~\ref{fig:all_evol} the evolution with time of the Specific Star
Formation Rate (SSFR), the ratio of  star formation rate
to  stellar mass.  Among the different definitions of the SFR in
Illustris we chose the one compatible with our choice of stellar
mass, i.e., the sum of the star formation rates of all gas cells within
twice the stellar half mass radius of the subhalo.  We fit the
following function to the SSFR evolution:
\begin{equation}
  \frac{M_{\rm{ssfr}}}{M_{\rm{acc}}}(t) = \begin{cases} 
    \alpha_{\rm{ssfr}} t + c_{\rm{ssfr}}, & \mbox{if } t < t_{\rm{ssfr}} \\ 
    \beta_{\rm{ssfr}} t + c'_{\rm{ssfr}}, & \mbox{if } t_{\rm{ssfr}} < t < t'_{\rm{ssfr}} \\
    \gamma_{\rm{ssfr}} t + c''_{\rm{ssfr}}, & \mbox{if } t > t'_{\rm{ssfr}}.     \end{cases}
  \label{equ:broken_line_ssfr}
\end{equation}
where $c' = (\alpha - \beta) t_{\rm{ssfr}} + c$ and $c'' = (\beta -
\gamma)t'_{\rm{ssfr}} + (\alpha - \beta)t_{\rm{ssfr}} + c$.

The evolution of the SSFR shows a clear transition between galaxies
that are on average star forming (SSFR $\sim 0.2\,\rm{Gyr}^{-1}$), and
quenched (SSFR $\sim 0.01\,\rm{Gyr}^{-1}$), which happens mostly
between $t_{\rm{ssfr}} = (1.21 \pm 0.06)\,\rm{Gyr}^{-1}$, and
$t'_{\rm{ssfr}} = (3.32 \pm 0.14)\,\rm{Gyr}^{-1}$.  The three
different phases are marked by a slope change in the SSFR evolution,
with a slower evolution before $t_{\rm{ssfr}}$, and after
$t'_{\rm{ssfr}}$, than in the transition phase.  This transition,
happening a few Gyr after the beginning of accretion, is consistent
with a slow-starvation delayed-quenching scenario of galaxy evolution
in clusters \citep{tollet2017}.

\begin{table}
\centering
\begin{tabular}{|c  | c c c| }
\multicolumn{4}{c}{All haloes} \\
\hline
		& $M_{\rm{DM}}$			& $M_*$				& SSFR					\\
\hline
$\alpha$& $-0.192 \pm 0.005$	& $0.058 \pm 0.001$	& $-0.086 \pm 0.018$	\\
$\beta$	& $-0.064 \pm 0.001$	& $0.003 \pm 0.001$	& $-0.366 \pm 0.011$	\\
$\gamma$& --					& --				& $-0.142 \pm 0.012$	\\
$c$		& $1.054 \pm 0.004$		& $1.002 \pm 0.001$	& $-0.674 \pm 0.012$	\\
$t$	& $1.72 \pm 0.05$		& $2.98 \pm 0.06$	& $1.21 \pm 0.06$		\\
$t'$	& --					& --				& $3.32 \pm 0.14$		\\
\hline
\end{tabular}
\caption{Best fit parameters of the evolution of the dark matter mass,
  stellar mass, and specific star formation rate (SSFR), as a function of
  time. The fits are performed on the median evolution over all the
  subhaloes of the three hosts. }
\label{tab:fit_evol}
\end{table}

In summary, looking at Fig.~\ref{fig:all_fit} it appears that during
the first infall, subhaloes lose around 40\% of their dark matter at
accretion, but continue to form stars. Compared to the dark matter
stripping, the star formation quenching is delayed, and only starts when
the subhaloes get closer to the centre of the cluster.  We note that
we perform the fits presented in Fig.~\ref{fig:all_fit} and
Table~\ref{tab:fit_evol} only up to $t - t_{\rm{acc}} = 6\,\rm{Gyr}$:
the small quantity of remaining subhaloes after that time makes the
signal much noisier. This also corresponds to the time scale
  at which, on average, subhaloes cross the host halo virial radius,
  where the influence of the host potential becomes much stronger.

\begin{figure} 
  \begin{center}
    \includegraphics{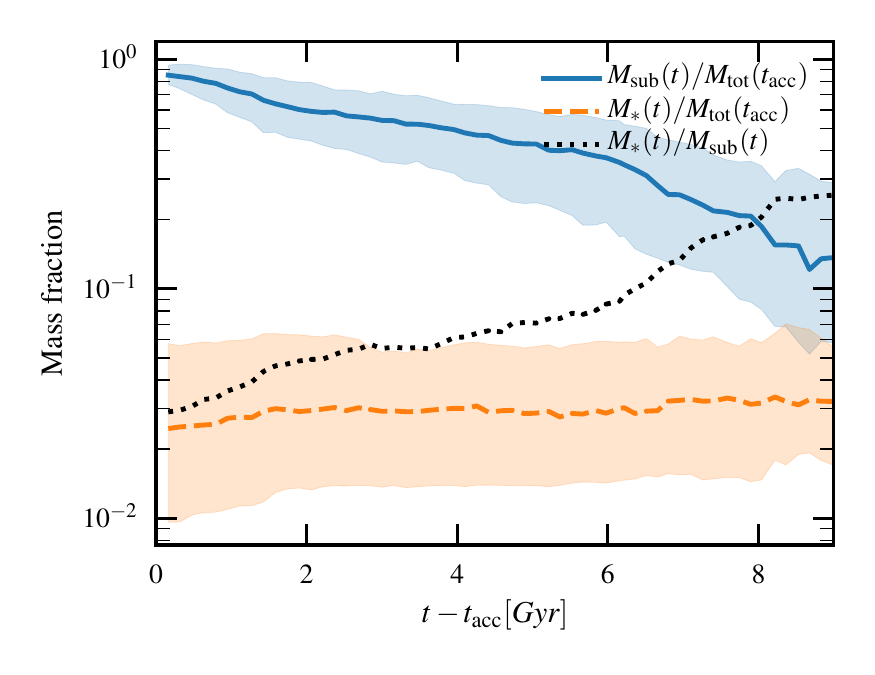}
    \caption{Evolution since the time of accretion of the dark matter
      (solid blue), and the stellar mass (dashed orange),
      normalized by the \textit{total} mass at accretion. The solid
      and dashed lines represent the median evolution. The filled
      surfaces show the $68\%$ credible interval. The
      evolution of the stellar to subhalo mass ratio is represented as the black
      dotted line. }
    \label{fig:Mdm_vs_Mstar}
  \end{center}
\end{figure}

In all our Figures so far, all mass evolution tracks are shown normalized
by the mass at accretion. To highlight the relative importance of the
dark matter and stellar mass loss, we show in
Fig.~\ref{fig:Mdm_vs_Mstar} the evolution of the dark matter mass
(blue solid line) and the stellar mass (orange dashed line), normalized
by the \textit{total} mass at accretion. This shows that 
the total mass evolution is dominated by the subhalo mass loss. It accounts for up to $90\%$ of the mass at accretion, while
the stellar mass increase represents only $\sim 2\%$ of the total mass
at accretion. However, due to these two effects, the proportion of
stellar and dark matter changes drastically during accretion: the
ratio of stellar to dark matter mass goes from 0.03 to 0.3 during
that time (black dotted line in Fig.~\ref{fig:Mdm_vs_Mstar}).

\section{Evolution of mass vs $R_{\mathrm{sat}}$}
\label{sec:m_vs_rsat}
	
To have a better representation of the relation between the
quenching/stripping, and the trajectory of the subhaloes, we look at the evolution of the dark matter and stellar mass, as well as the SSFR, as a function of distance to
the cluster centre. We keep the same subhalo selection as in
Sect.~\ref{sec:tevol}.

\begin{figure} 
   \begin{center}
     \includegraphics{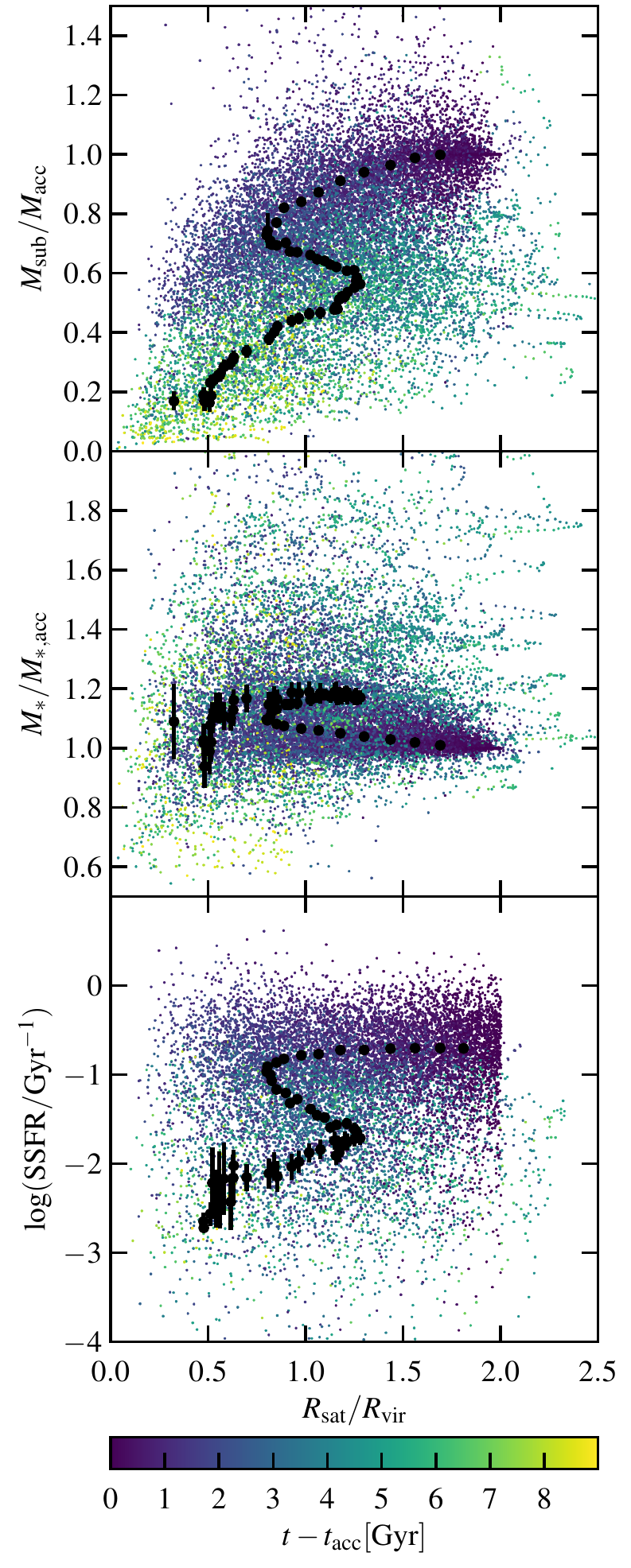}
     \caption{Position of the subhaloes on the $R_{\mathrm{sat}} -
       M_{\mathrm{sub}}/M_{\mathrm{acc}}$ (\textit{top panel}),
       $R_{\mathrm{sat}} - M_{*}/M_{*,\mathrm{acc}}$ (\textit{middle
         panel}), and $R_{\mathrm{sat}} - SSFR$ plane (\textit{bottom
         panel}) at each time step, starting when they are accreted at
       $2\times R_{200}$. The color of each point represents the time
       since accretion. The black dots represent the median values at
       each time step.}
     \label{fig:Rsat-all}
   \end{center}
\end{figure}

Starting with the dark matter mass, the top panel of Fig.~\ref{fig:Rsat-all} shows, at each time step
starting at their crossing of $2\times R_{200}$, the position of each
subhalo on the $R_{\mathrm{sat}} - M_{\mathrm{sub}}/M_{\mathrm{acc}}$
plane. 
We also show the
median value for all subhaloes at each time step as black dots,
with error bars corresponding to the standard error. The dark
matter mass appears to remain constant on average until subhaloes
reach $\sim 1.5\times R_{\rm{vir}}$. This would indicate that 
subhaloes only start to be affected by their host when they cross some
physical boundary of the halo. Such a physical boundary is now often
considered to be the splashback radius, which is defined as the radius
at which accreted matter reaches its first orbital apocenter after
turnaround \citep{more2015b, more2016, bush2017, baxter2017,
  diemer2017}. It has been measured to be located at
$\sim$\,1-2$\times R_{\rm{vir}}$, which is consistent with what we
observe.

The subhaloes then progressively lose their dark matter as they sink
towards the centre of the host. Looking at the mean evolution, $\sim 30\%$ of the dark matter mass is stripped at the
first pericentre,
$\sim 50\%$ after the first orbit, and up to $80\%$ for subhaloes that
spend $8-9\,\rm{Gyr}$ in their host.

We now look at the evolution of the stellar content of the
subhaloes. The middle panel of Fig.~\ref{fig:Rsat-all} shows the
evolution of the stellar mass as a function of the distance to the centre
of the host. The bottom panel of Fig.~\ref{fig:Rsat-all} shows the evolution of the SSFR. This
representation demonstrates more clearly how the delay in star formation
quenching relates to accretion in the host. On average, the satellite galaxies continue to form stars during the first infall (increase
in $M_*$, SSFR constant). The quenching process starts after the first passage at the pericentre.

In \citet{smith2016}, the authors study the
stripping of stellar and dark matter in galaxies during their infall
into simulated clusters, and found that $18\%$ of the galaxies were
undergoing important star formation during accretion, with an increase
in stellar mass higher than $15\%$. We test the influence of strongly
star forming galaxies by removing all galaxies that increase their
stellar mass by more that $50\%$ during their infall, which represent
$10\%$ of the total number of satellite galaxies in our
sample. Without them, the slope of the mean stellar mass increase
before the first passage at the pericentre is only slightly modified
(they reach $108\%$ of their initial stellar mass, instead of $109\%$
for the full sample). However, the sharp increase just after the pericentre
passage is dampened (the maximum stellar mass is $114\%$ of the initial
mass, compared to $120\%$ for the full sample at the same
moment). This could suggest that a small fraction of galaxies
experience a violent star formation burst, close to their passage at
the pericentre: observations of such star formation bursts in
infalling galaxies have been reported for example in
\citet{gavazzi2003}, who argue that it is caused by enhanced
ram-pressure during a high velocity infall.

\section{Summary and discussion}
\label{sec:summary}

 In this paper, we present a study of the evolution of satellite galaxies during their infall into the three most massive haloes of Illustris-1. We first measure the SHMR for
central, and satellite galaxies separately, and give a fitting function
for each case. We find that the SHMR for satellite galaxies is shifted
towards lower halo mass compared to central galaxies.  We find
  no dependence of this shift on the galaxy stellar mass, with a mean
  value $<\tau_{\rm{strip}}> = 0.75$, where $\tau_{\rm{strip}}$ is
  defined as the shift in subhalo mass at a given stellar mass (see
  equation \ref{equ:tau_strip}). This stripping factor is also in good
  agreement with the SHMR measured for central and satellite galaxies
  in \citet{rodriguez2013}.

We note that both for haloes and subhaloes, the SHMR we measure
differs at both mass ends from what is measured with abundance
matching \citep{moster2013,rodriguez2013}. There is in particular an
excess of massive galaxies that could be explained by an
underestimation of AGN feedback in the simulation that fails to
properly reduce star formation in massive haloes. It would therefore
be interesting to follow up on this work using the IllustrisTNG
simulation, which includes a new modeling of both stellar and AGN
feedback. In addition, the evolution of the SHMR that we measure between $z = 0$ and $z = 0.35$ is weaker than what was predicted in \citet{moster2013}. The larger volume of the IllustrisTNG-300 simulation set could be useful to study the redshift evolution of the SHMR for central and satellite galaxies in detail. 

To understand which mechanisms drive the shift of the SHMR for
satellite galaxies, we look at the time evolution of the stellar and
dark matter mass of the subhaloes during their infall. We find that
subhaloes lose a significant amount of their mass after their
accretion by the cluster (at a distance smaller than $\sim 1.5\times
R_{200}$).  As predicted by analytical models of tidal
  stripping, mass loss happens the fastest during the pericentric
  passages, with an average of $30\%$ of the initial mass lost during
  the first passage at the pericentre. This is in good agreement with
  the analysis presented in \citet{rhee2017}, where they found that
  subhaloes lose $20-30 \%$ of their initial mass during the first
  pericentric passage, the rest of the mass loss being attributed to
  subsequent passages at the pericentre, as well as to close encounters with
  other subhaloes. As an additional check, we bin subhaloes
  by their mass at infall, and measure the median evolution
  in each bin: from their first passage in the cluster, the most
  massive ones are found on tighter orbits, and lose their dark
  matter more quickly \citep[see also][]{xie&gao2015}. We find that
  during the first infall, less massive galaxies ($\log M_* < 10$)
  lose around 25\% of their initial mass, while the most massive
  galaxies ($\log M_* > 10$) around 40\%.  

The quenching of star formation is delayed compared to dark matter
stripping, and on average, galaxies stop forming new stars after their
first passage within the host core.  The median evolution of
  the SSFR suggests a slow quenching mechanism, with a quenching time
  $t_{\rm{Q}} \sim 2\,\rm{Gyr}$. This value is in very good agreement
  with the time scale estimated for instance in \citet{haines2015},
  and suggests that the dominant quenching mechanism is galaxy
  starvation.  To check a potential mass dependance of the star
  formation rate evolution, we split the galaxies by their
  stellar mass at infall, and measure the median evolution in each
  bin. We find two weak stellar mass dependencies: (i) galaxies in
  the lower mass bins ($7 < \log M_* < 9$) have a slightly steeper
  evolution than higher mass galaxies ($9 < \log M_* < 11$), leading
  to shorter quenching times ($t_{\rm{Q}} \sim 1.8\,\rm{Gyr}$ for low
  mass, and $t_{\rm{Q}} \sim 2.5\,\rm{Gyr}$ for higher mass); (ii)
  galaxies in the highest mass bin ($\log M_* > 11$) are already
  mostly quenched at infall. The latter effect could point towards an
  important role of pre-processing for high mass galaxies. However, this result should be
taken with care as our highest mass bin contains only $\sim
  30$ objects. Around 8\,Gyr after accretion, the average stellar
mass of satellite galaxies starts to decrease as well. This could
imply that stellar mass starts to be stripped as well when
subhaloes only have $\sim 30-40 \%$ of their remaining dark matter mass.
 This could also be an artifact due to the small number of
galaxies that have spent more than $7\,\rm{Gyr}$ in their hosts. However, this is in very good agreement with the
  evolution measured in \citet{smith2016}, where the stellar mass
  starts decreasing only after $70\%$ of the dark matter mass is
  stripped.

\begin{figure} 
   \begin{center}
     \includegraphics{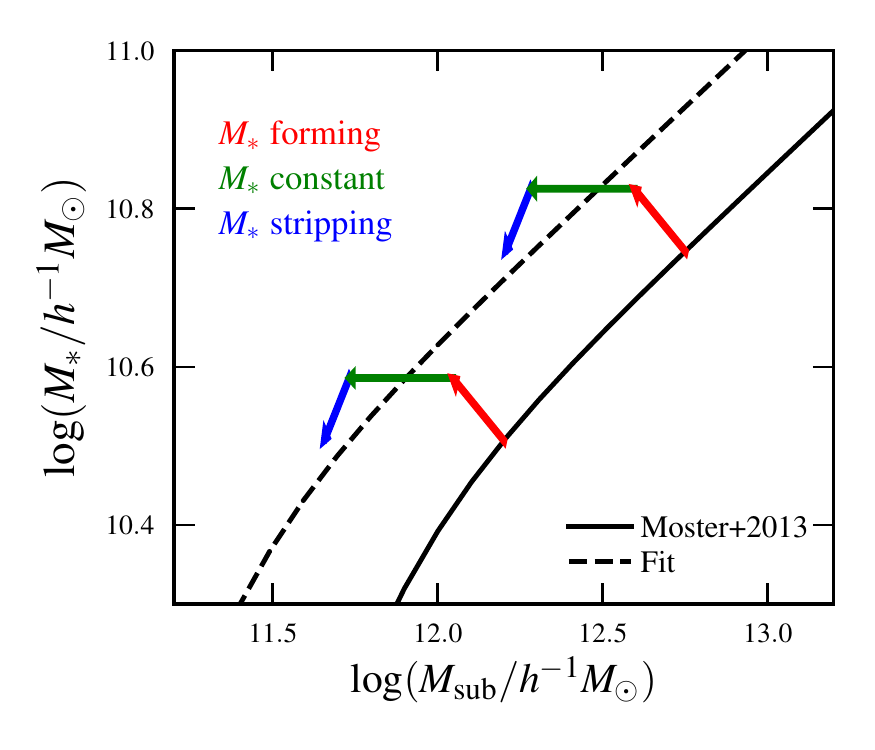}
     \caption{Summary of the subhalo SHMR evolution. The black solid
       line represents the SHMR for central galaxies at $z = 0$ from
       \citet{moster2013}, and the dashed line the fit from
       Sect.~\ref{sec:shmr_ill}. The arrows show the three different
       phases of the evolution: dark matter stripping + star formation
       in red, dark matter stripping only in green, and dark + stellar
       matter stripping in blue.}
     \label{fig:shmr_scenarios}
   \end{center}
 \end{figure}

We summarize the measured evolution of the SHMR of satellite galaxies in Fig. \ref{fig:shmr_scenarios}. The evolution can be
divided into three phases.
During the first one, which corresponds
roughly to the first infall, the galaxies lose on average $\sim 30\%$
of their dark matter mass at accretion, and continue to form stars,
reaching $\sim 120\%$ of their initial mass (red arrows in
Fig.~\ref{fig:shmr_scenarios}). Star formation is then quenched, and
the subhaloes continue to lose mass while the stellar mass remains
constant (green arrows). Finally, when only $30\%$ of the initial dark
matter mass remains, the average stellar mass starts to decrease as
well (blue arrows). We note that the evolution represented by the
arrows seems to predict a larger evolution of the SHMR than what we
measured (dashed line): this is simply due to the fact that not all
subhaloes follow this evolutionary path until the end, but are
distributed along the way.
We note that due to the limited size of the Illustris-1 simulation we did not investigate the dependence of the SHMR evolution on the host mass, but this could be tested with IllustrisTNG-300, which contains haloes with masses up to $10^{15}M_{\sun}$.

Although studies of simulations such as the one presented here allow
to disentangle the evolution of the dark component from the stellar one, it is
much more difficult in observational works. The only observable
that can potentially be obtained is the stripping factor $\tau_{\rm
  strip}$ which includes both the stripping of dark matter, and the
formation or stripping of stellar matter. However, if the true
evolution is similar to what simulations predict, dark matter
stripping should be the main contributor to the shift of the
SHMR.  
Figure~\ref{fig:shmr_scenarios} shows that the amplitude of the
subhalo mass evolution is $\sim 0.6$\,dex, while it is
only 0.1\,dex for the stellar mass. In any case, the stripping of subhaloes (or SHMR shift)
has not yet been measured with high confidence in observations. For instance,
weak gravitational lensing allows to statistically measure the total
mass of subhaloes, with a precision that is proportional to the area,
and the depth of the available observations. The advent of large
galaxy surveys such as DES or \emph{Euclid} in the future, could therefore
allow to put stronger observational constrains on the SHMR shift in
clusters.

Another aspect that remains observationally challenging is to estimate
the stage of accretion of galaxies in clusters, and even simply their
membership. A commonly used proxy is the projected distance to the
cluster centre, which is on average indeed correlated with the time
since accretion but adds noise coming from the shapes of the
individual orbits and from line-of-sight projections.
 Future data coming from upcoming ground-
  and space-based facilities will allow a better characterization
  of cluster membership and environment. It will also provide an increase of the
  statistical sample of several orders of magnitude \citep{sartoris16},
  and will allow us to shed more light on the dark and visible
  properties of satellite galaxies in clusters.

\section*{Acknowledgements}
We thank the Illustris collaboration for making their simulation data
publicly available.  MJ was supported by the Science and Technology
Facilities Council (grant number ST/L00075X/1).  CG acknowledges
support from the Italian Ministry of Foreign Affairs and International
Cooperation, Directorate General for Country Promotion (Project "Crack
the lens"); and the financial contribution from the agreement ASI
n.I/023/12/0 "Attivit\`a relative alla fase B2/C per la missione
Euclid". ML acknowledges CNRS and CNES for support.  \appendix

\section{Corner plots}
\label{sec:corner_plots}

\begin{figure*}
	\centering
	\includegraphics{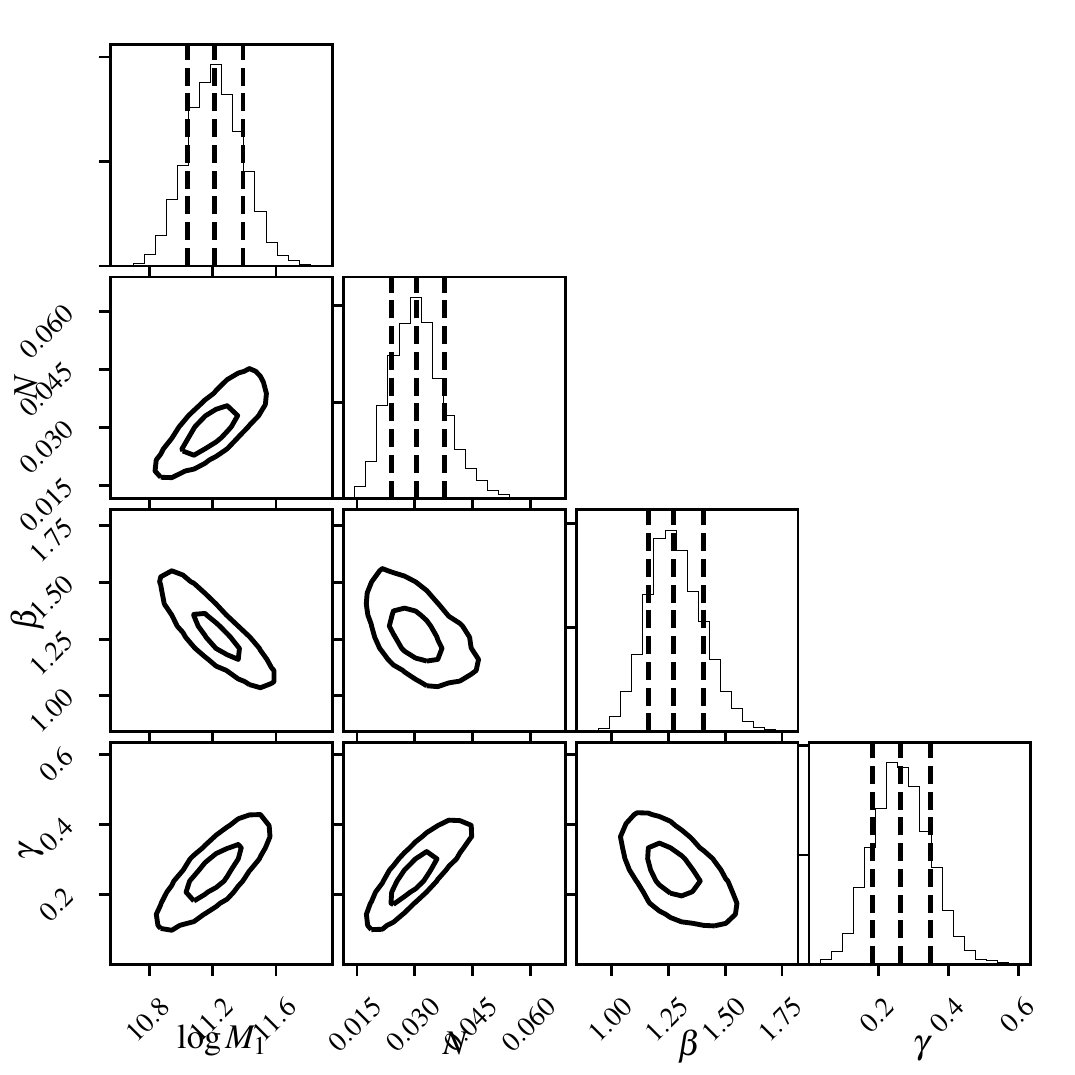}
	\caption{Joint 2-dimensional and marginalized 1-dimensional posterior probability distributions for the SHMR parameters $M_1$, $N$, $\beta$, and $\gamma$ fitted to central galaxies at $z = 0$.
          }
	\label{fig:cornerplot_haloes}
\end{figure*}

\begin{figure*}
	\centering
	\includegraphics{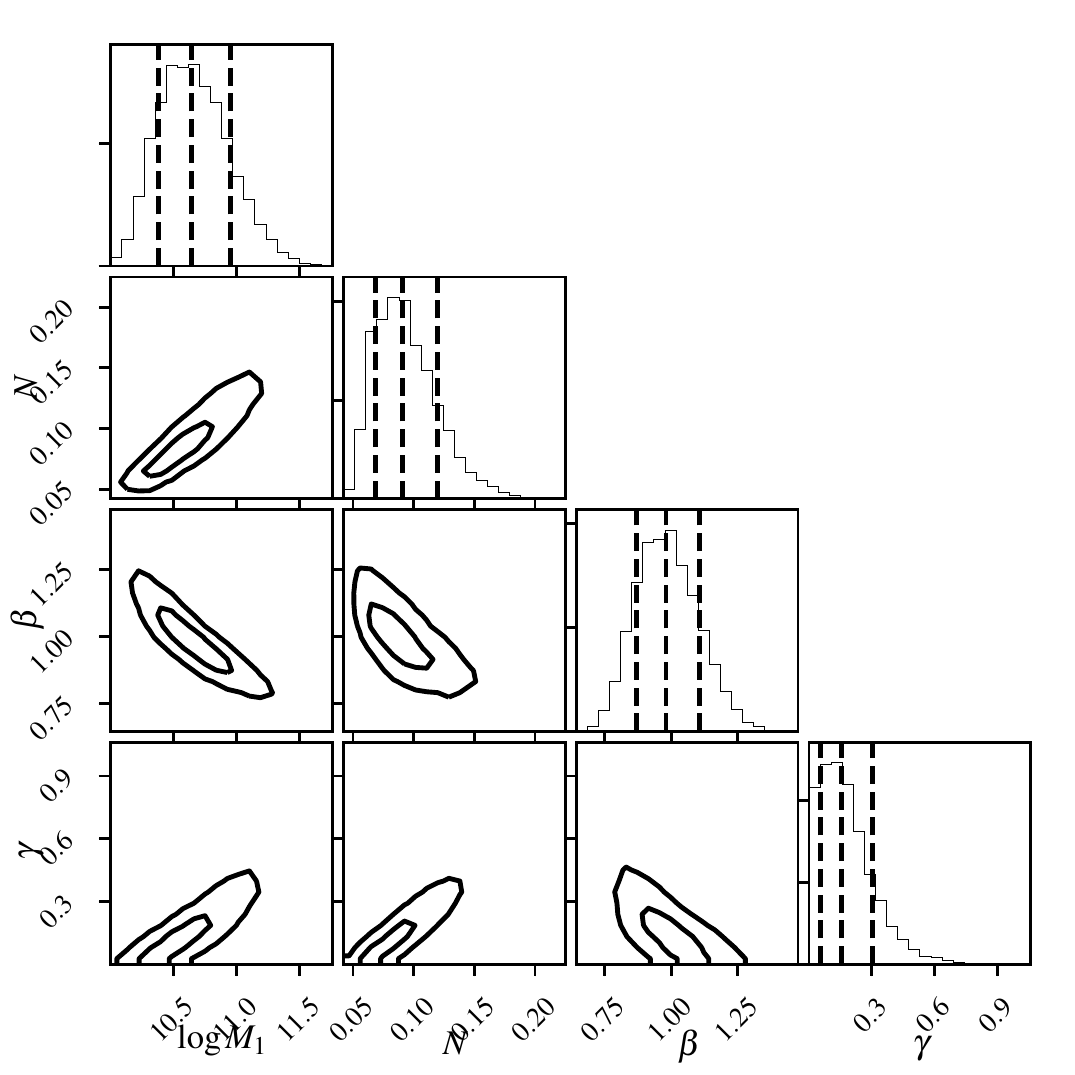}
	\caption{Same as Fig.~\ref{fig:cornerplot_haloes} but for satellite galaxies.
          }
	\label{fig:cornerplot_subs}
\end{figure*}

\bibliographystyle{mnras}
\bibliography{lensing.bib,globalbibs.bib}

% Don't change these lines
\bsp	% typesetting comment
\label{lastpage}
\end{document}